\documentclass[fleqn,usenatbib]{mnras}

\usepackage{newtxtext,newtxmath}

\usepackage[T1]{fontenc}
\usepackage{ae,aecompl}

\usepackage{graphicx}	
\usepackage{amsmath}	
\usepackage{amssymb}	

\defcitealias{podsiadlowski1995tzofate}{PCR95}

\newcommand{\Msun}{\ensuremath{\mathrm{M}_\odot}}
\newcommand{\Rsun}{\ensuremath{\mathrm{R}_\odot}}

\newcommand{\Msunpyr}{\ensuremath{\mathrm{\Msun~yr^{-1}}}}
\newcommand{\tzo}{T{\.Z}O}
\newcommand{\tzos}{T{\.Z}Os}

\title[Explosions of \tzos]{
Explosions of Thorne-\.Zytkow objects
}

\author[T. J. Moriya]{
Takashi J. Moriya\thanks{E-mail: takashi.moriya@nao.ac.jp}\thanks{NAOJ Fellow}
\\
Division of Theoretical Astronomy, National Astronomical Observatory of Japan, 
National Institutes of Natural Sciences, \\ 2-21-1 Osawa, Mitaka, Tokyo 181-8588, Japan
}

\date{Accepted 2018 January 5. Received 2018 January 4; in original form 2017 December 2}

\pubyear{2018}

\begin{document}
\label{firstpage}
\pagerange{\pageref{firstpage}--\pageref{lastpage}}
\maketitle

\begin{abstract}
We propose that massive Thorne-\.Zytkow objects can explode. A Thorne-\.Zytkow object is a theoretically predicted star that has a neutron core. When nuclear reactions supporting a massive Thorne-\.Zytkow object terminate, a strong accretion occurs towards the central neutron core. The accretion rate is large enough to sustain a super-Eddington accretion towards the neutron core. The neutron core may collapse to a black hole after a while. A strong large-scale outflow or a jet can be launched from the super-Eddington accretion disk and the collapsing Thorne-\.Zytkow object can be turned into an explosion. The ejecta have about 10~\Msun\ but the explosion energy depends on when the accretion is suppressed. We presume that the explosion energy could be as low as $\sim 10^{47}~\mathrm{erg}$ and such a low-energy explosion could be observed like a failed supernova. The maximum possible explosion energy is $\sim 10^{52}~\mathrm{erg}$ and such a high-energy explosion could be observed as an energetic Type~II supernova or a superluminous supernova. Explosions of Thorne-\.Zytkow objects may provide a new path to spread lithium and other heavy elements produced through the $irp$ process such as molybdenum in the Universe.
\end{abstract}

\begin{keywords}
accretion, accretion discs -- stars: neutron -- stars: peculiar -- supernovae: general -- supergiants  
\end{keywords}



\section{Introduction}
A Thorne-\.Zytkow object (\tzo) is a theoretically predicted star that has a degenerate neutron core. Its configuration is first studied by \citet{thorne1975tzo,thorne1977tzo}, although earlier studies speculate the existence of neutron-core supported stars \citep{gamow1937,landau1938tzo}. A stable \tzo\ configuration is found when its envelope mass is smaller than $\simeq 8~\Msun$ or larger than $\simeq 14~\Msun$ \citep{thorne1975tzo,thorne1977tzo,biehle1991tzo,biehle1994tzo,cannon1993tzo}. \tzos\ can be formed when a neutron star is swallowed by a companion star as a result of an unstable mass transfer \citep{taam1978tzo} or a supernova kick \citep{leonard1994tzo}.

The existence of \tzos\ is also questioned. For example, it is argued that accretion rates onto a neutron star during inspiral towards its companion core to form a \tzo\ would be high enough for the neutron star to collapse into a black hole \citep[e.g.,][]{brown1995nsacc,chevalier1996accretion,fryer1998hestargrb}. The black-hole formation could even trigger an explosion during the inspiral and \citet{chevalier2012common} argues that such an explosion during the inspiral could be observed as a Type~IIn supernova or a superluminous supernova because of a dense circumstellar medium formed during the inspiral. In this study, we assume that neutron stars avoid these possibilities during the inspiral and \tzos\ are successfully formed.

Massive ($\gtrsim 14~\Msun$) \tzos\ support their envelopes through nuclear reactions called $irp$ (interrupted rapid proton) process near the neutron cores \citep{cannon1993tzo,biehle1991tzo,tout2014tzo}. As a result, \tzos\ can be observed as red giants/supergiants with peculiar chemical abundances having, e.g., enhanced molybdenum and rubidium. It is also suggested that lithium abundance can be enhanced in \tzos\ \citep[PCR95 hereafter]{podsiadlowski1995tzofate}. A red supergiant star HV2112 is suggested to be a massive \tzo\ candidate because of its chemical anomalies that match the theoretical predictions \citep{levesque2014tzo}, although other possibilities are also suggested (\citealt{sabach2015soker}; \citealt{maccarone2016demink}, but see also \citealt{worley2016tzo}).

If the nuclear reactions are terminated in a massive \tzo\ because of exhaustion of $irp$ process seed elements or mass loss that put the \tzo\ below the mass limit ($\simeq 14~\Msun$) required to sustain the $irp$-process reactions, the \tzo\ configuration is no longer stable. \citetalias{podsiadlowski1995tzofate} find that such a \tzo\ triggers a strong accretion towards the central neutron core that may end up with a black hole formation. Here, we look into the \tzo\ collapse and suggest that a massive \tzo\ may explode in the end.

\section{Collapse}
After the termination of the nuclear burning at the neutron-core surface, matters around the surface keep contracting and the temperature becomes high enough for neutrino emission to dominate energy loss ($\gtrsim 2.5\times 10^{9}~\mathrm{K}$, \citetalias{podsiadlowski1995tzofate}). Thanks to the efficient neutrino cooling, the accretion towards the neutron core is not limited by the Eddington rate and a super-Eddington accretion can be immediately achieved. The subsequent accretion rate is therefore determined by a free-fall accretion rate.

\begin{figure}
	\includegraphics[width=\columnwidth]{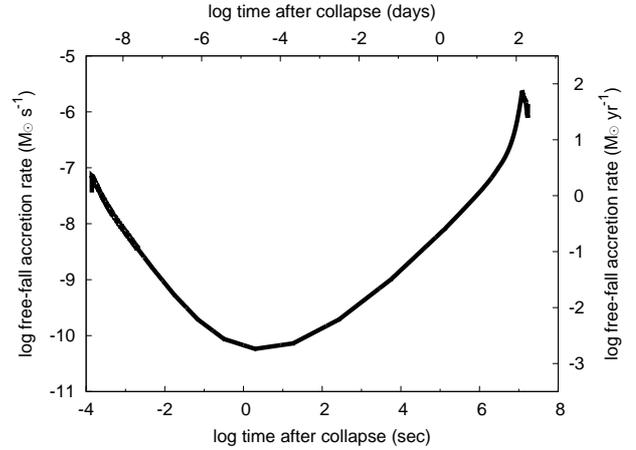}
    \caption{
    Free-fall accretion rate of a 16~\Msun\ \tzo\ with a 1~\Msun\ neutron core presented in \citet{biehle1991tzo}.
    }
    \label{fig:accretion}
\end{figure}

Fig.~\ref{fig:accretion} shows a free-fall accretion rate of a 16~\Msun\ \tzo\ with a 1~\Msun\ neutron core presented in \citet{biehle1991tzo}. Its radius is 1100~\Rsun. The accretion rate is kept above $\sim 10^{-3}~\Msunpyr$ which is required to trap photons in an accreting flow and keep a super-Eddington accretion \citep{chevalier1989nsacc}. Therefore, the accretion rate remains to be super-Eddington and follows the free-fall accretion rate in Fig.~\ref{fig:accretion}.

The free-fall accretion is suppressed when the centrifugal force starts to be dominant. \citetalias{podsiadlowski1995tzofate} suggest that \tzos\ rotate rigidly because of efficient convections in their envelopes. They estimate that a typical \tzo\ angular velocity is $\sim 3\times 10^{-9}~\mathrm{s^{-1}}$. In this case, the Keplarian rotation is achieved near the neutron core surface ($\sim 10~\mathrm{km}$) in a few days after the collapse. At this point, $\sim 10^{-3}~\Msun$ has been accreted on the neutron core and it has not collapsed to a black hole yet. A thick accretion disk would be formed around the central neutron core. The subsequent evolution depends on an uncertain viscosity in the accretion disk. The viscous timescale of the thick disk is $\sim (\alpha \omega_k)^{-1}$ where $\alpha$ is the viscosity parameter and $\omega_k$ is the Keplerian angular velocity \citep[e.g.,][]{kumar2008viscous}. Because $t_\mathrm{ff}\sim \omega_k^{-1}$, where $t_\mathrm{ff}$ is a free-fall time, the viscous timescale is $\sim \alpha^{-1} t_\mathrm{ff}$. Assuming a typical $\alpha \sim 0.1$, the viscous timescale is $\sim 10$ times longer than the free-fall timescale. Therefore, the accretion rate is reduced by a factor of $\sim 10$ compared to the free-fall accretion rate in Fig.~\ref{fig:accretion} after a few days since the collapse. However, the accretion rate is still much larger than $\sim 10^{-3}~\Msunpyr$ required to keep the super-Eddington accretion. The accretion disk around the neutron core may trigger large-scale outflows at this stage \citep[e.g.,][]{ohsuga2007nsacc}.

If the super-Eddington accretion continues, about $1~\Msun$ is accreted on the neutron star at a few months (free-fall, when the initial angular velocity is small) or a few years (viscous, as discussed above) after the collapse. At this moment, the central neutron core collapses to a black hole. The subsequent accretion towards the black hole is easily kept super-Eddington because photons are swallowed by the black hole. If a disk is formed, the disk accretion towards the central black hole is high enough to be radiatively inefficient and geometrically thick and, therefore, the super-Eddington accretion continues. Such an accretion disk can drive large-scale outflows \citep[e.g.,][]{kohri2005acc,dexter2013kasen} or jets \citep[e.g.,][]{blandford1977bz}.

\section{Explosion}
Because $\sim 10~\Msun$ typically exists at $\sim 1000~\Rsun$ in massive \tzos\ \citep{thorne1975tzo,thorne1977tzo,biehle1991tzo,biehle1994tzo,cannon1993tzo}, their binding energies are of the order of $10^{47}~\mathrm{erg}$. Indeed, the binding energy of the 16~\Msun\ \tzo\ in Fig.~\ref{fig:accretion} is $5\times 10^{47}~\mathrm{erg}$. As discussed in the previous section, a super-Eddington accretion disk is likely formed around the central compact object during the collapse of a \tzo. The accretion disk can launch large scale outflows and/or jets. Thanks to the small binding energy of \tzos, outflows and/or jets can easily push back the collapsing \tzos\ and make them explode. Observational properties of exploded \tzos\ strongly depend on the injected energy from the accretion disk and several possibilities can be considered as outcomes of \tzo\ explosions.

If the accretion is suppressed shortly after the binding energy ($\sim 10^{47}~\mathrm{erg}$) is released, the explosion energy is likely $\sim 10^{47}~\mathrm{erg}$ or less. The explosion may be observed as a low energy Type~IIP supernova. The plateau bolometric luminosity and duration are estimated to be $\sim 2\times 10^{38}~\mathrm{erg~s^{-1}}$ ($-7.1~\mathrm{mag}$) and $\sim 900~\mathrm{days}$, respectively, if we adopt a formalism of \citet{kasen2009woosley}. Because this luminosity is comparable to the progenitor luminosity, a \tzo\ may look as if it disappeared without an explosion. The explosion energy is similar to those of so-called `failed' supernovae \citep{nadezhin1980nlosse,lovegrove2013woosley} and a low-energy \tzo\ explosion may be confused with them. If a \tzo\ happens to be surrounded by a dense circumstellar medium, the ejecta may interact with the dense circumstellar media and it may be observed as low-luminosity Type~IIn supernovae like SN~2008S \citep[and references therein]{botticella2009sn2008s}.

If the accretion is not suppressed and the disk wind continues to blow, the \tzo\ explosion energy can be as high as, or even higher than, that of supernovae. Assuming a canonical energy conversion efficiency of accretions to outflows ($\sim 10^{-3}$), the explosion energy can be of the order of $\sim 10^{52}~\mathrm{erg}$ at most. Such an explosion is observed as an energetic Type~II supernova like OGLE-2014-SN-073 \citep{terreran2017ogle14-073} or an accretion-powered hydrogen-rich superluminous supernova \citep{dexter2013kasen} like SN~2008es \citep{gezari2009sn2008es,miller2009sn2008es,inserra2016hrichslsn}.

If a strong collimated jet is launched following a black hole formation, an exploding \tzo\ may be observed as an extremely long gamma-ray burst like Swift~1644+57 as proposed by \citet{quataert2012kasen}. A superluminous supernova like transients may also appear as a result of such a jet if it can penetrate the envelope \citep[e.g.,][]{nakauchi2013ultraslsn}.

\citetalias{podsiadlowski1995tzofate} estimate that the birth rate of \tzos\ in our Galaxy is around $2 \times 10^{-4}~\mathrm{yr^{-1}}$. The rate of \tzo\ explosions is likely to be similar to the birth rate, i.e., $\sim 10^{-4}~\mathrm{yr^{-1}}$ in our Galaxy. Given the Galactic supernova rate of $\sim 1~\mathrm{yr^{-1}}$ \citep{li2011snrate}, we expect one \tzo\ explosion in 10,000 supernovae. The rate is comparable to that of superluminous supernovae \citep[e.g.,][]{quimby2013slsnrate} and the future or even current transient surveys would be capable of finding them. \tzo\ explosions can occur in a wide range of metallicity and we do not expect a metallicity dependence in \tzo\ explosion sites. 

Because \tzo\ explosions can have an ejecta mass of $\sim 10~\Msun$ and an explosion energy that is similar to supernovae, distinguishing \tzo\ explosions from supernovae could be a challenge. The most notable difference between \tzo\ explosions and supernovae is the lack of a metal core in \tzo\ explosions. Supernovae show strong metal lines from their progenitors' core in late phases but \tzo\ explosions would not show such strong metal lines because of the lack of the metal core. Also, no radioactive elements are synthesized in \tzo\ explosions and they could simply disappear when a recombination wave in  hydrogen-rich ejecta reaches at the center if the accretion power is suppressed in late phases. The peculiar abundances predicted in massive \tzos\ may also be a clue to identify \tzo\ explosions.

The lack or small number of \tzo\ candidates discovered so far may imply that lifetimes of \tzos\ are rather short. The short \tzo\ lifetimes can be caused by strong mass loss that put \tzos\ below the mass limit to cease the $irp$ reactions early \citep[cf.][]{cannon1993tzo}. If mass loss is the major cause to terminate $irp$ process in \tzos, explosion properties of \tzos\ could be rather uniform because explosions are always triggered when the mass of \tzos\ goes below a certain mass limit. Especially, the ejecta mass of \tzo\ explosions could always correspond to the mass limit ($\simeq 14~\Msun$).

If massive \tzos\ explode, a large amount of lithium and other peculiar elements that result from $irp$ process would be spread into interstellar media through \tzos. Therefore \tzo\ explosions could be important in the chemical evolution in the Universe, especially for the uncertain origin of $rp$ process elements. Further studies are required to find the possible effect of \tzo\ explosions in chemical evolution.

\section{Summary}
We have shown that \tzos\ could be in a family of explodable stars. The termination of  nuclear reactions supporting a massive \tzo\ leads to a super-Eddington accretion towards the central neutron core. The free-fall accretion rate is alway above $\sim 10^{-3}~\Msunpyr$ and a super-Eddington accretion continues. An accretion disk could be formed around the neutron core after a while. Then, the accretion time is determined by the viscous timescale, not the free-fall timescale, but the accretion rate is estimated to be large enough to keep a super-Eddington accretion even after the accretion time is determined by the viscous timescale. The accretion disk can blow large-scale outflows or jets that trigger an explosion of a collapsing \tzo. The central core could collapse to a black hole at some point.

The ejecta mass of a \tzo\ explosion is $\sim 10~\Msun$ but the explosion energy depends strongly on how much energy the accretion disk can provide to the collapsing \tzo. A typical binding energy of \tzos\ is $\sim 10^{47}~\mathrm{erg}$ and the accretion may be prevented when $\sim 10^{47}~\mathrm{erg}$ is provided to the ejecta. The explosion energy in this case would be similar to those of `failed' supernovae and exploding \tzos\ may have similar properties to them \citep[e.g.,][]{lovegrove2013woosley}. We estimate that the maximum possible explosion energy of a \tzo\ explosion is $\sim 10^{52}~\mathrm{erg}$. Such a \tzo\ explosion would be observed as an energetic Type~II supernova or a hydrogen-rich superluminous supernova. If a jet is launched, a very long gamma-ray burst may also be accompanied.
If mass loss putting massive \tzos\ below the stable mass limit is a common trigger of \tzo\ explosions, \tzo\ explosion properties may be uniform. Especially, the ejecta mass could always correspond to the mass limit of $\simeq 14~\Msun$.

The rate of \tzo\ explosions is estimated to be $\sim 10^{-4}~\mathrm{yr^{-1}}$. \tzos\ are known to synthesize a large amount of lithium and $rp$ process elements. Because \tzos\ explode, they may be an important source of heavy elements produced by an exotic $rp$ process as well as lithium in the Universe.

\section*{Acknowledgements}
I would like to thank the anonymous referee for valuable comments that improved this work. 
The author is supported by the Grants-in-Aid for Scientific Research of the Japan Society for the Promotion of Science (16H07413, 17H02864).




\bibliographystyle{mnras}
\bibliography{reference} 

\bsp	
\label{lastpage}
\end{document}